\documentclass[a4paper,10pt]{article}
\usepackage[utf8x]{inputenc}
\usepackage{jheppub}
\usepackage{booktabs}
\usepackage{blindtext}
\usepackage{amsmath}
\usepackage{bbm}
\usepackage{placeins}
\usepackage{subfig}
\usepackage{hepunits}
\usepackage{cancel}

\begin{document}

\title{The ALPs from the Top: Searching for long lived axion-like particles from exotic top decays}
\author[a]{Adrian Carmona,}
\author[b]{Fatemeh Elahi,}
\author[b]{Christiane Scherb,}
\author[b]{and Pedro Schwaller}
\affiliation[a]{CAFPE and Departamento de F\'isica Teórica y del Cosmos,\\ Universidad de Granada, E18071 Granada, Spain}
\affiliation[b]{PRISMA$^+$ Cluster of Excellence \& Mainz Institute for Theoretical Physics,\\ Johannes Gutenberg University, 55099 Mainz, Germany}
\emailAdd{adrian@ugr.es}\emailAdd{felahi@uni-mainz.de}\emailAdd{cscherb@uni-mainz.de}\emailAdd{pedro.schwaller@uni-mainz.de}
\abstract{
We propose a search for long lived axion-like particles (ALPs) in exotic top decays. Flavour-violating ALPs appear as low energy effective theories for various new physics scenarios such as t-channel dark sectors or Froggatt-Nielsen models. In this case the top quark may decay to an ALP and an up- or charm-quark. For masses in the few GeV range, the ALP is long lived across most of the viable parameter space, suggesting a dedicated search. We propose to search for these long lived ALPs in $t\bar{t}$ events, using one top quark as a trigger. We focus on ALPs decaying in the hadronic calorimeter, and show that the ratio of energy deposits in the electromagnetic and hadronic calorimeters as well as track vetoes can efficiently suppress Standard Model backgrounds. Our proposed search can probe exotic top branching ratios smaller than $10^{-4}$ with a conservative strategy at the upcoming LHC run, and potentially below the $10^{-7}$ level with more advanced methods. Finally we also show that measurements of single top production probe these branching ratios in the very short and very long lifetime limit at the $10^{-3}$ level.
}

\preprint{MITP/22-017}
\maketitle
\section{Introduction}
More than 10 years after the start of the LHC, the search for new physics continues. With increasing luminosity, the hunt for light but very weakly coupled new particles becomes more and more feasible. One prime example are axions or axion-like particles (ALPs), which are pseudo-scalar fields whose mass is protected from large corrections by an approximate shift symmetry. They are of great interest as possible solutions to the strong CP problem~\cite{Peccei:1977hh, Peccei:1977ur, Weinberg:1977ma, Wilczek:1977pj}, but are also predicted as lightest new degrees of freedom in many new physics scenarios such as composite models~\cite{Gripaios:2009pe, Ferretti:2013kya, Belyaev:2016ftv, Gripaios:2016mmi, Cacciapaglia:2019bqz, BuarqueFranzosi:2021kky}, strongly coupled dark sectors~\cite{Strassler:2006im, Han:2007ae, Bai:2013xga, Schwaller:2015gea, Renner:2018fhh, Cheng:2019yai, Cheng:2021kjg}, supersymmetric models~\cite{Bellazzini:2017neg,Gorbunov:2000ht} or in models with horizontal symmetries~\cite{Davidson:1981zd, Wilczek:1982rv, Reiss:1982sq, Berezhiani:1990jj, Berezhiani:1990wn, Feng:1997tn, Albrecht:2010xh, Bauer:2016rxs, Ema:2016ops, Calibbi:2016hwq, Ema:2018abj, Heikinheimo:2018luc, Bonnefoy:2019lsn}.

Given the current lack of experimental guidance for choosing a new physics scenario, a promising approach is to systematically parameterise the ALP couplings to standard model (SM) particles using effective lagrangians~\cite{Georgi:1986df, Choi:1986zw, Chala:2020wvs, Bauer:2020jbp, Bonilla:2021ufe}. 
ALPs with flavour violating couplings to one type of SM fermions are predicted in various new physics models such as t-channel dark sectors~\cite{Renner:2018fhh} or Froggat-Nielsen models of flavour~\cite{Froggatt:1978nt,Alanne:2018fns} where only one type of right-handed (RH) quarks have non-zero charges. The case where the ALP couples dominantly to RH up-type quarks was  studied in~\cite{Carmona:2021seb} (see also \cite{Bauer:2021mvw}), and it was shown that this model is poorly constrained in particular for ALP masses above the charm quark threshold. In this mass range the flavour violating coupling to the top quark offers new possibilities for experimental probes, using both precision top quark physics as well as new search strategies where the top quark is used as a trigger object. In this work we will explore both avenues. 

The flavour violating coupling of the top quark to a lighter quark and an ALP allows for exotic top decays, as well as direct production of the ALP in association with a top quark. The ALP mainly decays to hadrons, either promptly or with a long lifetime. Therefore it can easily contribute to single top events. In the first part of our work, we show that precision measurements of the single top cross section are able to probe this new physics scenario. We perform a recast of existing single top searches and obtain new constraints on the parameter space of the ALP for both prompt ALP decays to jets and for detector stable ALPs. 

Furthermore we propose a new strategy to search for ALPs in events containing pairs of top quarks. Thanks to the humongous cross section of $t\bar{t}$ events at hadron colliders, even a small branching ratio of the top into an ALP and a light jet will lead to a large rate of top plus ALP events. ALP decays are easily distinguishable from SM jets when they are displaced from the primary vertex, which is possible for ALPs close to the lower end of the allowed mass range. For decays happening in the hadronic calorimeter, one expects only a small energy deposit in the electromagnetic calorimeter as well as fewer tracks associated with the jet. This can be exploited to suppress the backgrounds by several orders of magnitude, and thus our proposed search will be sensitive to very small exotic top branching ratios. 

As usual, our paper starts with an Introduction, followed by an overview of the charming ALP model and its interactions. In Section~\ref{sec:constraints} the bounds from a recast of existing searches and constraints are presented. The newly proposed search for long lived ALPs produced in association with a top quark is introduced in Section~\ref{sec:search}, before concluding. Projections for the high luminosity LHC as well as further details on the simulations are available in the Appendix. 

\section{Charming ALPs and exotic top decays}

Similarly to Ref.~\cite{Carmona:2021seb}, we focus on scenarios where ALPs only interact with up-type quarks at tree-level, that we have dubbed charming ALPs. In this case, the relevant EFT reads 
\begin{align}
\label{eqn:EFT}
\mathcal{L}=\frac{1}{2}(\partial_{\mu}a)(\partial^{\mu}a)-\frac{m_a^2}{2}a^2+\frac{\partial_{\mu}a}{f_a}(c_{u_R})_{ij}\bar{u}_{Ri}\gamma^{\mu} u_{Rj}\,,
\end{align}
with 
\begin{align}
c_{u_R} = \begin{pmatrix}
c_{11}&c_{12}&c_{13}\\
c_{21}&c_{22}&c_{23}\\
c_{31}&c_{32}&c_{33}\\
\end{pmatrix},
\end{align}
a hermitian matrix. Such EFTs will be generated at tree-level by UV completions involving dark QCD-like sectors with scalar mediators or some models of flavour à la Froggatt-Nielsen, see e.g. the discussion in Ref.~\cite{Carmona:2021seb}. 

\begin{figure}
    \centering
       \includegraphics[width=0.49\linewidth]{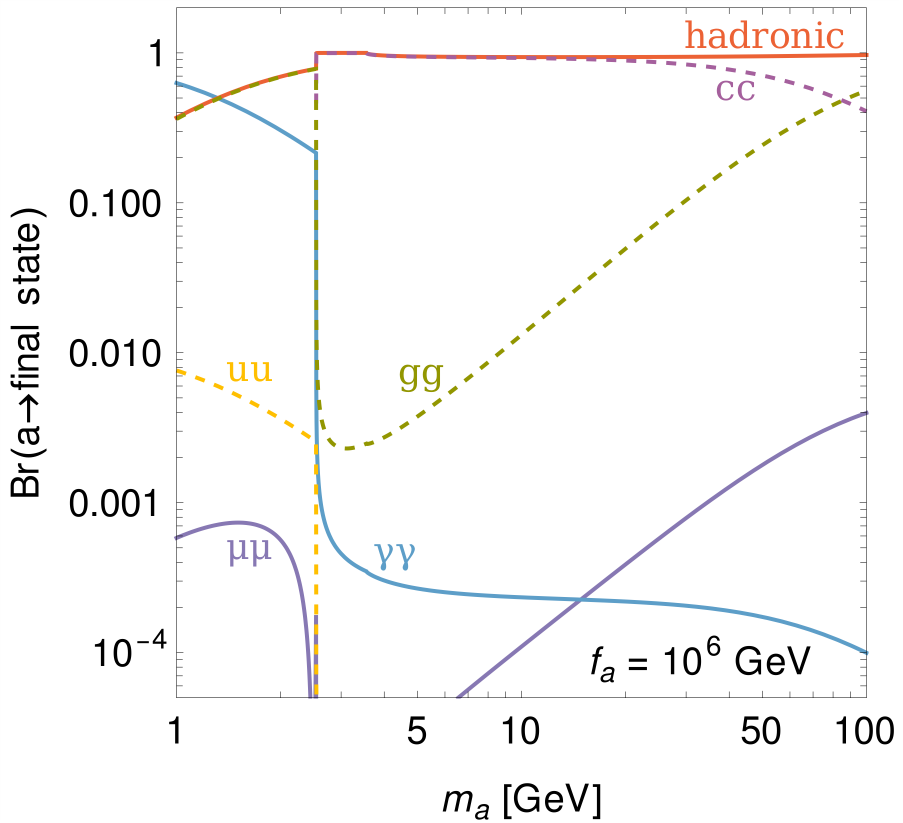}
        \includegraphics[width=0.49\linewidth]{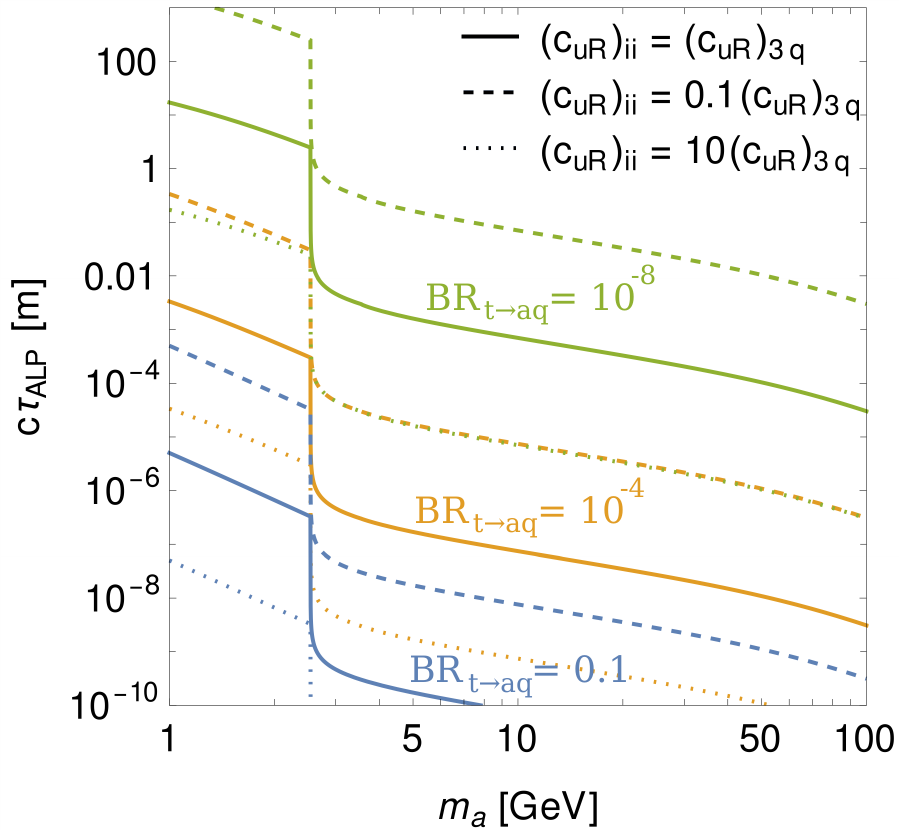}
    \caption{Left: ALP decay branching ratios as a function of the ALP mass $m_a$ for $f_a=10^6$~GeV. The dashed lines show the contributions to the hadronic channel.
    Right: Lifetime of the ALP as a function of the ALP mass. The blue, orange and green lines show $\mathrm{Br}(t\to a q) = 10^{-1}, 10^{-4}$ and $10^{-8}$, respectively. Solid, dashed and dotted lines refer to $(c_{u_R})_{ii}/(c_{u_R})_{3q} = 1, 0.1$ and $10$.}
    \label{fig:lifetimes}
\end{figure}

Of course ALP couplings to vector bosons and to other SM fermions (down-type quarks and leptons) will be generated radiatively via top loops and from the renormalization group equations (RGEs)~\cite{Chala:2020wvs, Bauer:2020jbp, Bonilla:2021ufe}. 
While these operators are suppressed relative to the tree level interactions of Eq.~(\ref{eqn:EFT}), they induce decays that can be relevant in some regions of the parameter space where the hadronic channels are kinematically inaccessible. This can be seen in Fig.~\ref{fig:lifetimes}, where we plot the different ALP branching ratios as a function of the ALP mass $m_a$ for $(c_{u_R})_{12}=0=(c_{u_R})_{21}$ and $(c_{u_R})_{ij}=1$ otherwise, as well as  $f_a=10^6$~GeV. Loop-generated decays like $a\to \mu^+ \mu^-$, $a\to gg$ or $a\to \gamma \gamma$ have been computed using the expressions present in Ref.~\cite{Carmona:2021seb} and we have used the quark-hadron duality~\cite{Poggio:1975af, Shifman:2000jv}  to compute the inclusive hadronic decay rate.  When $a\to \bar{c}c$ is not kinematically allowed, $a\to gg$ tends to dominate the ALP branching ratio. This channel also dominates for large enough values of the ALP mass, since the loop generated vector boson decays grow as $m_a^3$ while the fermionic decay widths are linear in the ALP mass.  

One should note that the RGE-induced decays into two fermions are logarithmically sensitive to the scale of the matching $\Lambda\sim f_a$, so smaller values of $f_a$ will reduce their relative impact. Small enough values of $(c_{u_R})_{12}$ and $(c_{u_R})_{21}$ are required in order to evade constraints from $\bar{D}^0-D^0$ mixing. For values of $m_a$ below $1$~GeV, one would need to use chiral perturbation theory instead of perturbative QCD but we focus here on the case $m_a\gtrsim 1$~GeV that is much less constrained by current searches (see results of Ref.~\cite{Carmona:2021seb}).

Light particles that mainly decay to hadrons are difficult to find at hadron colliders such as the LHC, due to the large amount of hadronic background events. Two features of our scenario will make such a search possible however, namely the presence of flavour violating decays in the up-quark sector and the possibly long lifetime of the ALP. Since neutral meson mixing  requires $(c_{u_R})_{12}$ and $(c_{u_R})_{21}$ to be extraordinarily small, a novel and interesting way of searching for ALPs at colliders is to concentrate on flavour-violating top decays involving long-lived ALPs. As can be seen from the right panel of Fig.~\ref{fig:lifetimes}, the ALPs decay length can easily reach the typical length scales of LHC detectors for masses in the $1~\rm{GeV}-10~\rm{GeV}$ range. To simplify the parameter space, the couplings are chosen as
$(c_{u_R})_{ii}\equiv (c_{u_R})_{11}=(c_{u_R})_{22}=(c_{u_R})_{33}$ and  $(c_{u_R})_{3q}\equiv (c_{u_R})_{13}=(c_{u_R})_{23}=(c_{u_R})_{31}=(c_{u_R})_{32}$. While the ALPs lifetime is mainly set by the diagonal coupling $(c_{u_R})_{ii}$, the exotic top decay depends on the off diagonal couplings $(c_{u_R})_{3q}$ via 
\begin{align}
    \mathrm{Br}(t\to a q_i) &= \frac{N_c}{96\pi}\frac{|(c_{u_R})_{3i}|^2}{f_a^2}\frac{m_a^2}{m_t}\left(\frac{\left(m_i^2-m_t^2\right)^2}{m_a^2}- \left(m_t^2+m_i^2\right)\right)\nonumber\\
    &\times\sqrt{\left(1-\frac{\left(m_a+m_i\right)^2}{m_t^2}\right)\left(1-\frac{\left(m_a-m_i\right)^2}{m_t^2}\right)}\times \frac{1}{\Gamma_t},\quad q_i=u,c\,,
    \label{eq:Br}
\end{align}
with $\Gamma_t$ the total top decay width.

Since all branching ratios just depend on the ratio $c_{u_R}/f_a$ and the exotic decays 
only involve off-diagonal couplings, we can use $\mathrm{Br}(t\to aq)$ and the ratio $(c_{u_R})_{ii}/(c_{u_R})_{3q}$ as free parameters. In particular, we represent $c\tau_{\rm ALP}$ as a function of $m_a$  for $\mathrm{Br}(t\to aq)=10^{-1},10^{-4}$ and $10^{-8}$ as well as hierarchies of diagonal versus non-diagonal couplings of $(c_{u_R})_{ii}/(c_{u_R})_{3q}=0.1,1,10.$ One can then readily see that the `natural' mass region to find long-lived ALPs without resorting to tiny values of $\mathrm{Br}(t\to aq)$ is $m_a\sim 1-10$~GeV. Moreover, since the ALP decay width is dominated by decay modes involving diagonal couplings, we can trade $\{\mathrm{Br}(t\to aq),(c_{u_R})_{ii}/(c_{u_R})_{3q})\}$ for $\{\mathrm{Br}(t\to aq), c\tau_{\rm ALP}\}$. We will use this last set of ALP lifetime and exotic top branching ratio as our independent model parameters for our phenomenological studies.  
Note that a change in $f_a$ can be absorbed in a redefinition of the couplings (up to small logarithmic corrections to the branching ratios). We can therefore fix it to an arbitrary scale which we choose to be $f_a = 10^6~{\rm GeV}$.

\section{Experimental constraints}
\label{sec:constraints}

\subsection{Model independent limits on exotic top decays}

Even though the top quark was discovered more than two decades ago, measuring its decay width is still a challenging process. Direct measurements of the top decay width, which avoid model-dependent assumptions, have large uncertainties: $0.6 < \Gamma_t < 2.5\ \text{GeV}$ at $95\%$ C.L. ~\cite{ CMS:2016hdd,ATLAS:2017vgz}. This is mainly because of the low experimental resolution to reconstruct the jet-related properties (e.g, jet reconstruction, jet energy resolution, jet energy scale, jet vertex fraction)~\cite{ATLAS:2011lgt,CDF:2013xca,CMS:2016lmd,ATLAS:2020cli}. New methods that use combinations of resonant and non-resonant cross-sections to extract a model independent top quark decay width measurement have been proposed~\cite{Baskakov:2017jhb,Baskakov:2018huw,Baskakov:2019bjb}, which can reduce the uncertainties significantly: $\Gamma_t = 1.28\pm 0.30 \ \text{GeV}$ ~\cite{Herwig:2019obz}. However, $O(10 \%)$ uncertainties still allow for large new physics contributions. Indirect measurements of the top decay width have less uncertainties, but they are done under certain SM assumptions~\cite{D0:2012hgn,CMS:2014mxl}. Hence, they are not applicable when searching for new physics in rare top decays.

Nonetheless, flavour-changing neutral current (FCNC) decays involving the top have been under extensive experimental scrutiny. In particular, the $t q X$ coupling with $q = u, c$ and $X = h/Z/\gamma/ g$ is carefully studied~\cite{ATLAS:2018zsq,CMS:2017wcz,CMS:2017twu,CMS:2015kek,ATLAS:2019mke,ATLAS:2018jqi,CMS:2016obj,CMS:2017bhz,ATLAS:2015iqc,CMS:2016uzc}. The SM prediction for FCNC top decay is diminutive because of loop and CKM suppression: $\text{Br} ( t \to q X) \ll 10^{-10}$~\cite{Balaji:2020qjg}. Therefore, new physics contributions can feasibly be persued (e.g, Refs. ~\cite{Andrea:2011ws,Kamenik:2011nb,Banerjee:2018fsx,Ebadi:2018ueq,Castro:2020sba}). Due to the resemblance between a light quark jet and a $b$-jet, however, FCNC top coupling searches are usually focused on exotic top quark productions\,\footnote{In cases where exotic top decays have been studied, the properties of $X$ (mass, decay products, etc) are used to tease out the signal~\cite{ATLAS:2018zsq,CMS:2017wcz,CMS:2017twu,ATLAS:2018jqi,CMS:2016obj,CMS:2017bhz}.}, in the form of a single top plus X searches. In the following, we will narrow our attention to top + jets and single top production to find the current bounds. 
That is because if the ALP decays at the scale of the detector length, then the final state becomes top + jets, while if the ALP leaves the detector before decaying, then the signature becomes a single top.    

\subsection{Recast of searches for exotic top decays}
\label{sec:recast}
\begin{figure}[t]
    \centering
    \includegraphics[width=0.67\linewidth]{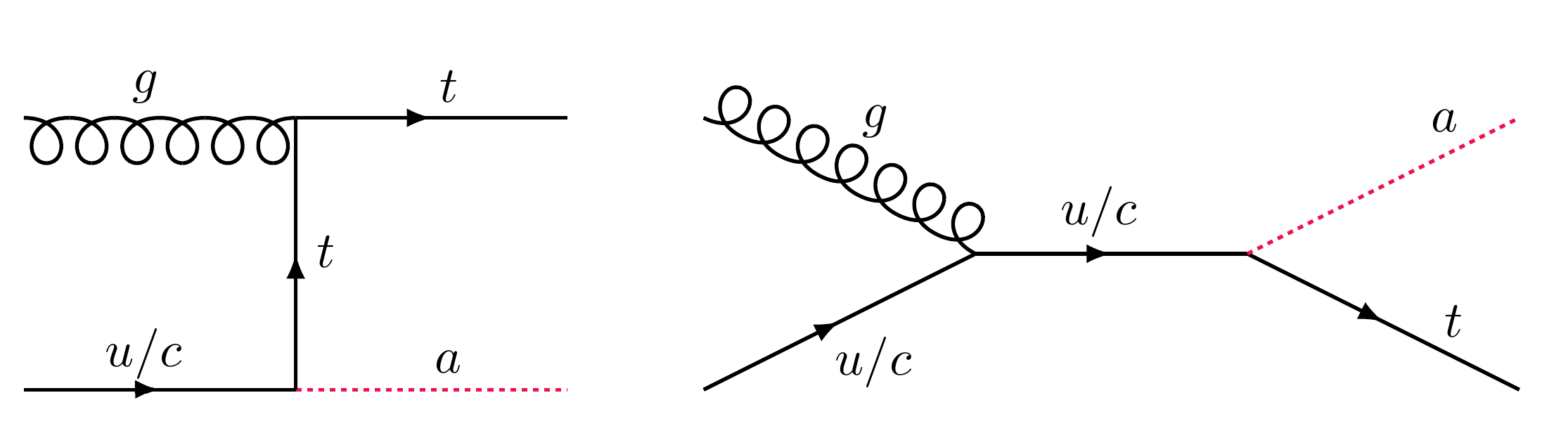}
    \caption{The Feynman diagrams representing top + ALP production at the LHC.}
    \label{fig:feyta}
\end{figure}

One of the dominant processes at the LHC involving the charming ALP is its production in association with a single top. The main diagrams for top + ALP production are shown in Fig.~\ref{fig:feyta}. Knowing that for the mass range of our interest the ALP mainly decays hadronically, top + jets searches can impose some constraints on ALP couplings. The CMS experiment has conducted a search in the top+jet channel probing the anomalous $tqg$ coupling~\cite{CMS:2016uzc}. Specifically, they looked for a leptonic top in association with one or two jets, where at least one of them fails the b-tagging secondary vertex algorithm. This algorithm selects jets with $0.01 \ \text{cm} < r < 2.5 \  \text{cm}$, where $r$ is the radial distance between the secondary vertex and the primary vertex~\cite{CMS:2017wtu}. Since in this search, they want a jet that fails the b-tagging algorithm, and gluon and light quark jets tend to have prompt vertices, it is clear that $r < 0.01$ cm is considered in their search. However, it is unclear whether $ r > 2.5$ cm is considered in their search. To stay conservative, we will assume that jets with $ 2.5 \ \text{cm} < r < 2 \text{m}$ are not rejected\footnote{ If the ALP has not decayed by the hadronic calorimeter ($r \simeq  2$ m), it cannot be detected as a normal jet.}, and we recast the results accordingly. Given that the upper limit on the cross section of new physics contributing to $pp \to t+j$ is $ \sigma_{tj} \simeq  0.29\  \text{pb}$ at $\sqrt{s} = 13 \ \text{TeV}$~\cite{Goldouzian:2016mrt}, an upper limit on $\left(c_{u_R}\right)_{3q}/f_a$ with $q= u,c$ can be found using MadGraph5 \cite{Alwall:2014hca}. Then, using Eq.~(\ref{eq:Br}), this can be converted into an upper limit on $\text{Br}(t\to aq)$. In deriving this limit, we have to take into account the probability that the ALP decays such that it is (most likely) accepted by the search. For prompt decays with $ r < 0.01$, the efficiency factor is 
\begin{align}
\int_0^{10^{-4}\,\rm{m}}  (\gamma c\tau_{\rm ALP})^{-1} e^{ -\frac{ct}{\gamma {c\tau_{\rm ALP}}}}  d(ct)\,,
\end{align} 
where $\gamma = p_T/ m_a$ is the boost factor along the transverse direction. The MC generated events were weighted according to the boost factor.
Similarly, for ALPs that decay in the range $ 2.5 \ \text{cm} < r < 2 \ \text{m}$, we include an efficiency factor 
\begin{align}\int_{2.5 \times 10^{-2}\, \rm{m}}^{2\,\rm{m}}  (\gamma c\tau_{\rm ALP})^{-1} e^{ -\frac{ct}{\gamma {c\tau_{\rm ALP}}}}  d(ct).  
\end{align} 
The dark green regions in Fig.~\ref{fig:exclusion} represent the constraints coming from the top+jet search at CMS. The dashed line is the constraint for $ \text{Br}(t \to au)$ and the solid line is for $\text{Br}(t \to ac)$. In this work, we are interested in studying a long-lived ALP. Hence, the constraint coming from $ r< 0.01 \ \text{cm}$ is not visible in the Fig.~\ref{fig:exclusion}, except at the bottom right corner of the plot for $m_a = 10 \ \text{GeV}$.  In general, larger boost factors (smaller $m_a$), push the constraints to lower $c\tau$.

If the ALP is stable on the scale of the detectors, it will appear as missing energy. In this case, measurements of single top production rates impose some constraints on the couplings of the ALP. Single top production in the SM is suppressed by the $b$ quark PDF and therefore relatively small. The ATLAS experiment searched for top FCNC with gluon mediator in the single top channel~\cite{ATLAS:2015iqc}, and reported an upper limit in the cross section ($\sigma_t \lesssim 0.10 \ \text{pb}$ at $\sqrt{s} = 13 \ \text{TeV}$)~\cite{Goldouzian:2016mrt}. In their analysis they require exactly one jet, one lepton, and missing energy, and they use Multivariate Analysis to find their limit. One of the variables they used as an input is the transverse mass $m_{T_{\ell \nu}}$, which should have an upper limit of $m_W$ in the case of true single top production. In the case of top + ALP where ALP is another source of missing energy, however, $m_T$ should have a different distribution. Nonetheless, we recast their limits to stay conservative regarding the potential power of the LHC in constraining ALP couplings in this channel.  In this case, to take into account the probability that the ALP does not decay on the scale of the detector $(ct \geq 10 \ \text{m})$ one has to introduce an efficiency factor of $e^{ -\frac{10\, \rm{m} }{\gamma {c\tau_{\rm ALP}}}}$.
%
 The light green regions in Fig.~\ref{fig:exclusion} demonstrate the constraint that the ATLAS search imposes on our model. The dashed line is the limit for $ \text{Br}(t \to au)$, and the solid line is for $ \text{Br}(t \to ac)$. 
 
Similarly, searches for single top + transverse missing energy (MET) without FCNCs can be used to constraint the parameter space. However, single top + MET searches are typically performed for dark matter candidates with masses $\mathcal{O}(100)~\text{GeV}$ \cite{ATLAS:2018cjd,ATLAS:2020yzc,CMS:2019zzl,CMS:2018gbj} and require at least 200~GeV of MET. Instead in our scenario the typical amount of MET is $\lesssim m_t/2$ if the ALP escapes undetected, and therefore most events would fail the experimental selection.

\section{Search strategies and LHC prospects for top decays to long lived particles}
\label{sec:search}
\subsection{Signal properties}
\begin{figure}[t]
    \centering
    \includegraphics[width=0.52\linewidth]{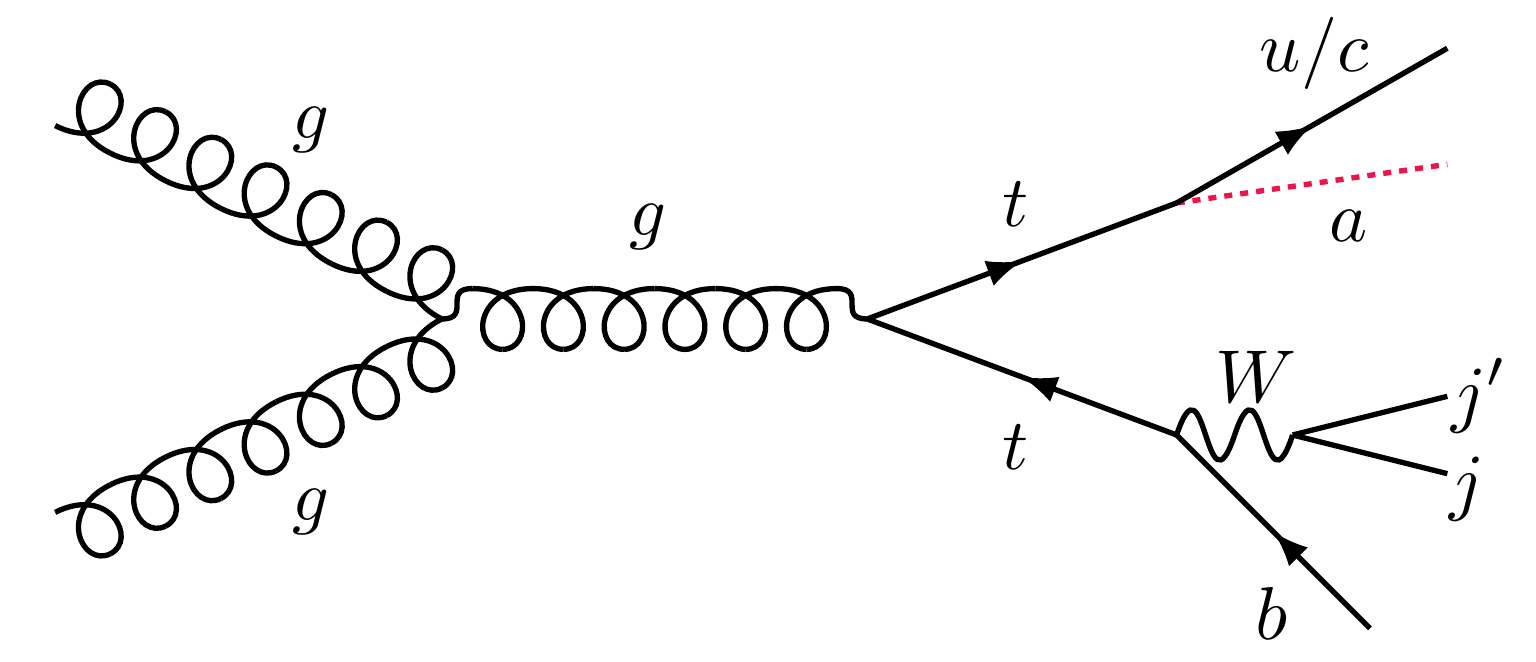}
    \caption{The Feynman diagram for the signal: $t\bar t$ production, where one of the tops decay to $q= u,c$ and an ALP.}
    \label{fig:feytt}
\end{figure}

For the search proposed here, we focus on ALP production via flavour-violating top decays. More precisely, we consider top-pair production where one of the tops decays via its main SM decay mode to $Wb$ and the other to an ALP and either an up or charm quark, see Fig.~\ref{fig:feytt}. Consequently, the signal production cross section is 
\begin{align}
\sigma_{\rm signal} = \sigma_{t\bar{t}}\times \mathrm{Br}(t\to W b)\times \mathrm{Br}(t \to a q),    
\end{align}
with $\sigma_{t\bar{t}} \sim 830$~pb \cite{ATLAS:2020aln}, $\mathrm{Br}(t\to Wb)\sim 0.96$ \cite{Zyla:2020zbs} and $\mathrm{Br}(t\to a q)$ given in Eq.~(\ref{eq:Br}).
 For couplings $(c_{u_R})_{ij}$ of order one and $\frac{1}{f_a}\sim\mathcal{O}(10^{-9}-10^{-5})$~GeV$^{-1}$ light ALPs with $m_a \sim (1-10)$~GeV have lifetimes of order millimeter to 100~m, while having $\mathrm{Br}(t\to a q)\lesssim 10^{-3}$. For these intermediate lifetimes ALPs decay mostly in the hadronic calorimeter or the muon spectrometer. We should remind the reader that while the ALP decays to pairs of partons, it is highly boosted and decays displaced, so it will mainly be reconstructed as a single, narrow jet. In the following, we will distinguish two different cases: the case where the ALP decays at the outer edge of the electromagnetic calorimeter or inside the hadronic calorimeter and the case where it decays in the muon spectrometer.

An ALP decaying inside the hadronic calorimeter leads to a jet that deposits most of its energy in the hadronic calorimeter and thus, to a large value of the hadronic to electromagnetic energy ratio $E_{\rm had}/E_{\rm em}$. Since the ALP is neutral we expect no tracks associated with the jet from its decay. In addition to the displaced jet from the ALP, the signal  consists of one prompt light jet from the up or charm quark produced in the flavour violating decay $t\to a q$ ($q = u, c$) and one to three prompt jets, one of them being a $b$-jet, from the decay of the second top quark. The main background in this case is $t\bar{t}$, where a jet consisting of (anti-) protons, $\pi^\pm$ and/or $K^\pm$, but no photons, deposits the majority of its energy in the hadronic calorimeter and is thus reconstructed as a ``displaced" jet. However, such a jet will leave tracks, a feature we will use to distinguish signal and background.

On the other hand, if the ALP decays in the muon spectrometer, the signal consists of an event originating in the muon system with no associated tracks pointing to the primary vertex, as well as the same prompt jets as for decays in the hadronic calorimeter. Consequently, we expect $2-4$ ($2-5$) jets and a hit in the muon spectrometer without any associated tracks. We assume that this signal is background free.   

\subsection{Triggering and event selection}
First, we focus on ALP decays inside the hadronic calorimeter. Here, the signal consists of minimal three and maximal five (six) jets, one (two) of them being displaced: the decay products of a SM decay of a top, a prompt light jet from the flavour violating top decay and one (or rarely two) displaced jet(s) from the ALP decay. In general, one could reconstruct both top quark masses, one from the displaced jet and one additional jet, and the other from the remaining three jets, to reduce the background. However, we found that focussing on the displaced jet provides sufficient background suppression and such a reconstruction of the invariant top masses is not necessary. Note that here and in the following, we treat the top quark that decays to SM final states as a collider observable object, since the experimental collaborations have demonstrated that they can trigger on and identify top quark decays with high efficiency and accuracy. 
We therefore do not explicitly implement top-tagging, however we do demand that the jets from the top decays are reconstructed with large enough transverse momenta, so that we do not overestimate the sensitivity of the search. 

We therefore select events with $3-6$ ($3-5$) jets with $p_T > 40$~GeV and $|\eta| < 2.5$. 
To identify the displaced jet we follow the ATLAS Calorimeter Ratio trigger \cite{ATLAS:2013bsk} requirements. This trigger is taking advantage of the fact that the decay products of neutral particles decaying in the outer layers of the electromagnetic calorimeter or in the hadronic calorimeter deposit most of their energy in the hadronic calorimeter. The Calorimeter Ratio trigger requires a $\tau$-lepton like object with $E_T > 40$~GeV (which fits the jet originating from the ALP), with $\log_{10}\left(E_{\rm had}/E_{\rm em}\right) > 1.2$ and no tracks with $p_T > 1$~GeV in a ($0.2\times0.2$) region in ($\Delta\eta\times\Delta\phi$) around the jet direction.

\begin{figure}
    \centering
    \includegraphics[width=0.47\linewidth]{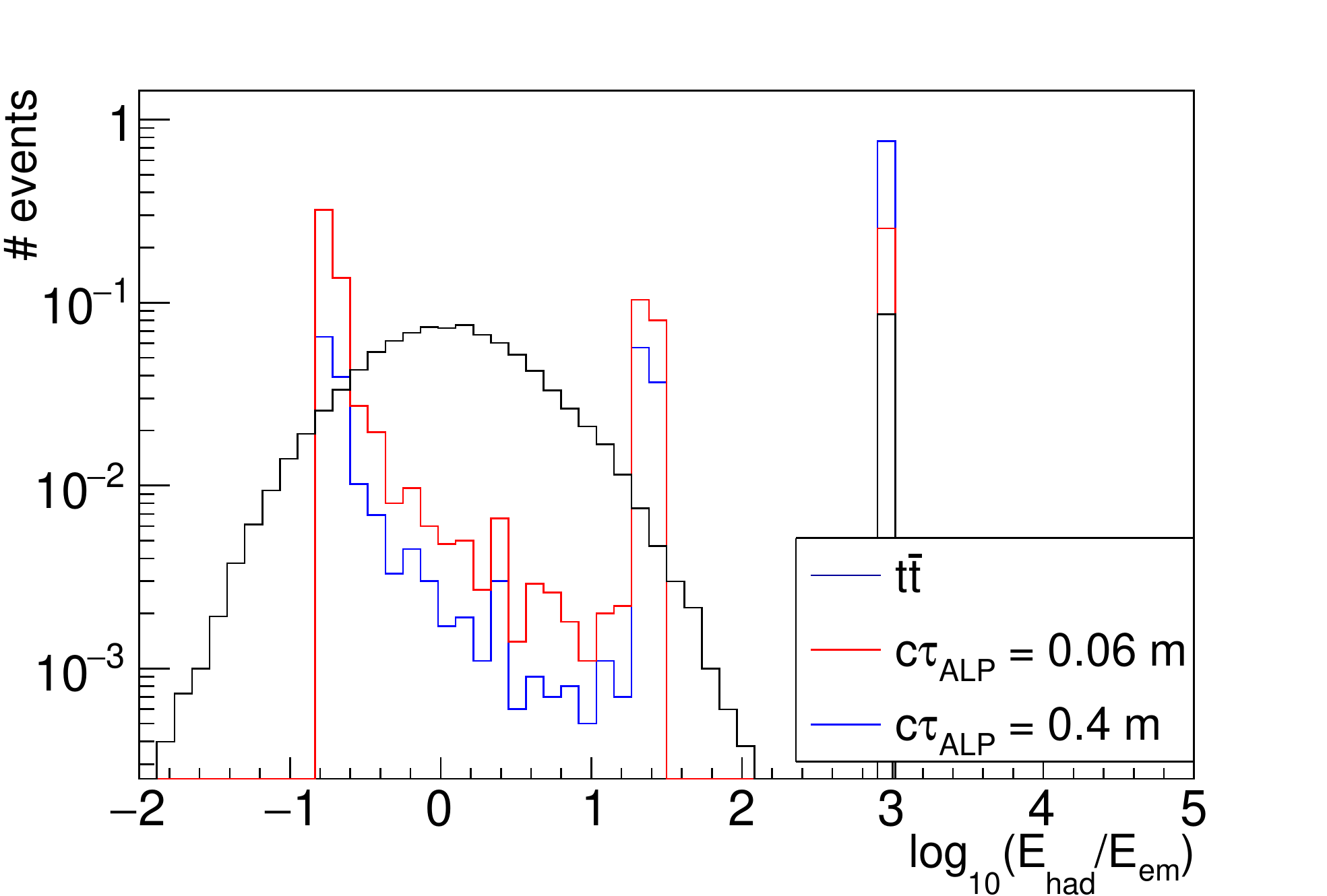}
    \quad
    \includegraphics[width=0.47\linewidth]{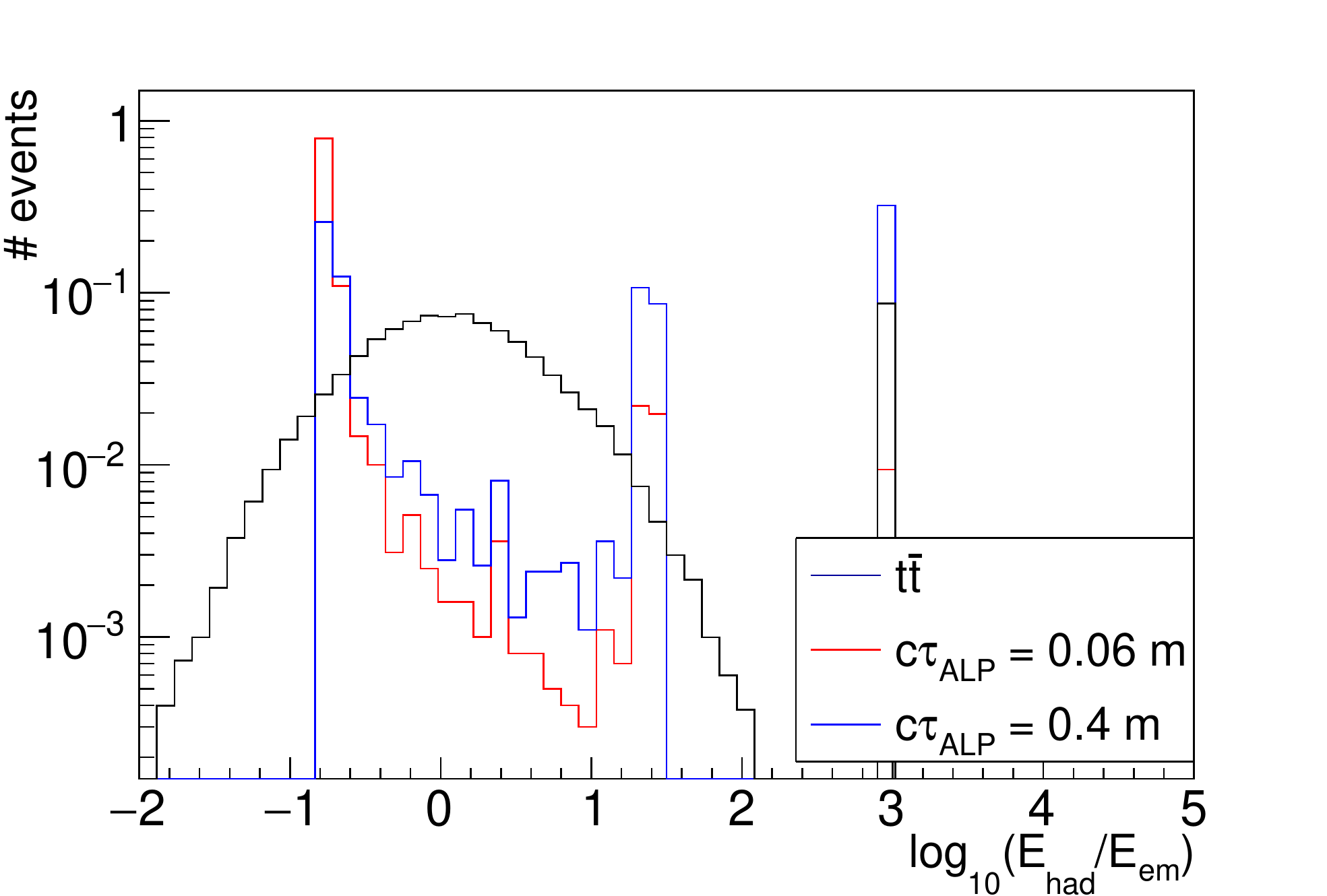}
    \caption{Calorimeter energy deposit ratio $\log_{10}\left(E_{\rm had}/E_{\rm em}\right)$ distribution for the $t\bar{t}$ background as well as for the signal with $c\tau_{\rm ALP}=0.06,\,0.4$\,m and $m_a = 2$~GeV (left) and $m_a = 10$~GeV (right).}
    \label{fig:Ehad_over_Eem}
\end{figure}

In Fig.~\ref{fig:Ehad_over_Eem} we show the $\log_{10}\left(E_{\rm had}/E_{\rm em}\right)$ distribution for the signal with $m_a = 2$~GeV (left) and $m_a = 10$~GeV (right) and ALP lifetimes $c\tau_{\rm ALP} = 0.06$~m and $0.4$~m, as well as for the $t\bar{t}$ background. We modified the FeynRules \cite{Christensen:2008py, Alloul:2013bka} implementation of the linear ALP EFT model \cite{Brivio:2017ije, Brivio:2017ALPEFT} to include the charming ALP couplings.  Signal events were generated with MadGraph5 \cite{Alwall:2014hca} with showering and hadronization done with Pythia8 \cite{Sjostrand:2014zea}. The energy deposit ratio $\log_{10}\left(E_{\rm had}/E_{\rm em}\right)$ for the signal was assigned according to Fig.~5b of \cite{ATLAS:2013bsk}. For background estimation we simulated $100000$ $t\bar{t}$ events with MadGraph5 \cite{Alwall:2014hca} with showering and hadronization done with Pythia8 \cite{Sjostrand:2014zea} and fast detector simulation carried out by Delphes \cite{deFavereau:2013fsa}. 

While the background in Fig.~\ref{fig:Ehad_over_Eem} is evenly distributed around $\log_{10}\left(E_{\rm had}/E_{\rm em}\right)\sim 0$, corresponding to an equal energy deposit in the hadronic and electromagnetic calorimeter, and has one peak in the overflow bin at $\log_{10}\left(E_{\rm had}/E_{\rm em}\right) = 3$, the signal has several peaks: The signal peak at  $\log_{10}\left(E_{\rm had}/E_{\rm em}\right)\sim -0.8$ corresponds to the ALPs that decay close to the interaction point. The second peak with $\log_{10}\left(E_{\rm had}/E_{\rm em}\right) \gtrsim 1.2$ is due to ALPs decaying in the outer layers of the electromagnetic calorimeter or inside the hadronic calorimeter. Thus, they only deposit a small amount of energy in the electromagnetic calorimeter. Note that this peak is higher for $c\tau_{\rm ALP} = 0.06$~m when $m_a =2$~GeV and for $c\tau_{\rm ALP} = 0.4$~m when $m_a = 10$~GeV, respectively. This is due to the fact that the ALP is less boosted for higher masses. Finally, the signal has a peak at $\log_{10}\left(E_{\rm had}/E_{\rm em}\right) = 3$, similar to the background, but it has a different origin than for the background. For the signal this peak shows the amount of ALPs decaying outside of the detector and thus, do not count into the actual signal, while for the $t\bar{t}$ background it shows jets with $E_{\rm em}= 0$ and therefore $\left(E_{\rm had}/E_{\rm em}\right) = \infty$, which is defined as $\left(E_{\rm had}/E_{\rm em}\right) \equiv 1000$ in the Delphes cards, leading to  $\log_{10}\left(E_{\rm had}/E_{\rm em}\right)\equiv 3$. As described above this is true for jets consisting of (anti-) protons, $\pi^\pm$ and/or $K^\pm$, but no photons. These jets are counted as signal. 

To further reduce the background of SM jets that appear displaced, we now consider the no track criterion of the Calorimeter Ratio trigger. 
\begin{figure}
    \centering
    \includegraphics[width=0.67\linewidth]{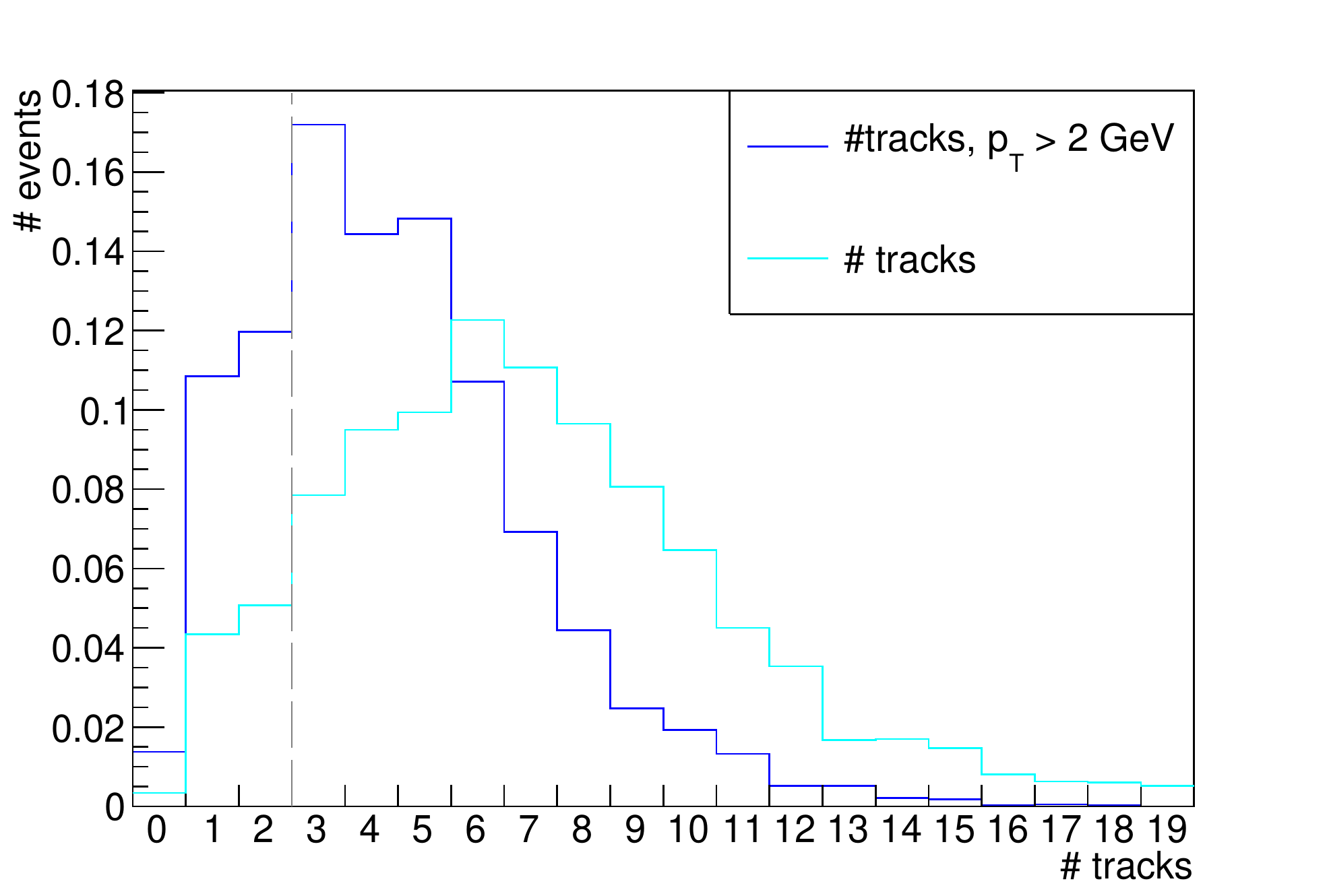}
    \caption{Number of tracks of jets originating from $t\bar{t}$ with $\log_{10}\left(E_{\rm had}/E_{\rm em}\right) > 1.2$. The dark blue line shows the number of tracks with $p_T > 2$~GeV, the lighter blue line the number of all tracks for such a jet.}
    \label{fig:number_tracks}
\end{figure}
In Fig.~\ref{fig:number_tracks} the number of tracks for background jets with $\log_{10}\left(E_{\rm had}/E_{\rm em}\right) > 1.2$ is shown. The light and dark blue lines correspond to all tracks and to tracks with $p_T > 2$~GeV. It can be seen that in both cases most jets have at least one track. At the level of our simulation, the signal events have no tracks pointing towards the decaying ALP. However in reality pile-up events could add tracks pointing in the direction of the displaced decaying ALP, and thus a very strict cut on the tracks could reduce the sensitivity. 
We therefore choose a less stringent cut on the number of tracks for background jets as the actual Calorimeter Ratio trigger, requiring that jets with $\log_{10}\left(E_{\rm had}/E_{\rm em}\right) > 1.2$ have less than two tracks with $p_T > 2$~GeV. This cut is indicated by the grey dashed line in Fig.~\ref{fig:number_tracks}. Even with this conservative cut most of the background will be removed.

In addition, we consider a search for ALPs decaying in the muon spectrometer. Here, we select events with $2-5$ prompt jets with $p_T > 40$~GeV and $|\eta| < 2.5$, while the ALP should decay inside the muon calorimeter ($4.3$~m $< L_{xy} < 10.7$~m) and fulfill $p_T > 25$~GeV and $|\eta| < 2.5$. 
\subsection{LHC sensitivity and prospects at future colliders}
  For two ALP masses $m_a = 2$~GeV and $m_a = 10$~GeV, we generated $10000$ signal events for various lifetimes $c\tau_{\rm ALP}= 0.001-100$~m with MadGraph5 and Pythia8, as before. We select events in Pythia8 with $3-6$ ($3-5$) jets, each with $p_T > 40$~GeV and $|\eta| < 2.5$. We demand that the jet from the ALP fulfils the $\log_{10}\left(E_{\rm had}/E_{\rm em}\right) > 1.2$ criterion of the Calorimeter Ratio trigger according to the energy deposit ratio as a function of the decay radius in Fig.~5b of \cite{ATLAS:2013bsk} and further demand that the ALP satisfies $p_T > 40$~GeV and $|\eta|< 2.5$. 
For background estimation, we select events with $3-6$ ($3-5$) jets with $p_T > 40$~GeV and $|\eta| < 2.5$ (from $100000$ $t\bar{t}$ events generated with MadGraph5, Pythia8 and Delphes). In addition we require that at least one of these jets has $\log_{10}\left(E_{\rm had}/E_{\rm em}\right) > 1.2$ and that this jet has no more than two tracks with $p_T > 2$~GeV. 

As experimental testing grounds we consider LHC with $\sqrt{s} = 13$~TeV and the expected total integrated luminosity after run 3 of $\mathcal{L}= 350$~fb$^{-1}$, as well as the high-luminosity phase of LHC (HL-LHC) with $\sqrt{s} = 14$~TeV and a total integrated luminosity $\mathcal{L}= 4000$~fb$^{-1}$.   In Tab.~\ref{table:cutflow} the cut flow (of the efficiencies) for the signal for $m_a = 2$~GeV and $m_a= 10$~GeV with $c\tau_{\rm ALP}= 0.1$~m and $\mathrm{Br}(t\to a q) = 0.001$, as well as for the background is shown for $\sqrt{s} = 13$~TeV and the expected total integrated luminosity $\mathcal{L}= 350$~fb$^{-1}$. 
\begin{table}[]
    \centering
    \begin{tabular}{l|c|c|c}
         & $\mathbf{m_a = 2}$~\textbf{GeV} & $\mathbf{m_a = 10}$~\textbf{GeV} & $\mathbf{t\bar{t}}$ \\
         \hline
        \textbf{total} & (1) $2.79\times10^{5}$ & (1) $2.79\times10^{5}$ & (1) $2.91\times 10^{8}$ \\
        \hline
        $\mathbf{3-6}$ \textbf{jets with}  &  & &\\
        \textbf{$\mathbf{p_T>40}$~GeV \&} $\mathbf{|\eta| < 2.5}$ & (0.8439) $2.35\times10^{5}$& (0.8414) $2.35\times10^{5}$ & (0.71801) $2.09\times10^{8}$ \\
        \hline
        \textbf{1 jet with} $\mathbf{log_{10}\left(\frac{E_{\rm had}}{E_{\rm em}}\right)> 1.2}$ & (0.1436) $4.00\times10^4$ & (0.0775) $2.16\times10^4$ &(0.01244) $3.61\times 10^{6}$\\
        \hline
         \textbf{displaced jet has $\mathbf{\leq 2}$ tracks}& (0.1436) $4.00\times10^4$ & (0.0775) $2.16\times10^4$ & (0.00022) $6.39\times 10^4$ \\
         \textbf{with $\mathbf{p_T > 2}$~GeV}& & & \\
    \end{tabular}
         \caption{Cut flow of the expected number of events for signal and background events for LHC run 3 with $\sqrt{s} = 13$~TeV and $\mathcal{L}= 350$~fb$^{-1}$. The values in brackets are the efficiencies after each cut. For the signal $c\tau_{\rm ALP} = 0.1$~m and $\mathrm{Br}(t\to a q) = 0.001$ was chosen.}
         \label{table:cutflow}   
\end{table}
It can be seen that already the cut of minimal three and maximal six jets with $p_T > 40$~GeV and $|\eta|<2.5$ reduces the background compared to the signal, however the cuts on $\log_{10}\left(E_{\rm had}/E_{\rm em}\right)$ and the number of tracks are significantly stronger and allow to clearly distinguish signal and background. Depending on the mass and lifetime of the ALP up to $\sim 15\%$ of the ALP signal passes these cuts, while each of them reduces the number of background events by about two orders of magnitude. In Tab.~\ref{table:cutflow2} in the appendix the same cutflow is shown for choosing events with three to five jets with $p_T > 40$~GeV and $|\eta|<2.5$. This reduces signal and background in a similar way and thus does not improve the signal to background ratio. Based on the above described selection criteria we perform a cut-and-count analysis, using $S/\sqrt{S+B}=2$ to find the expected $2\sigma$ exclusion region. Since we expect that the backgrounds can be further suppressed, we do not include systematic effects in our sensitivity estimate. 

\begin{figure}[th]
    \centering
    \includegraphics[width=0.63\linewidth]{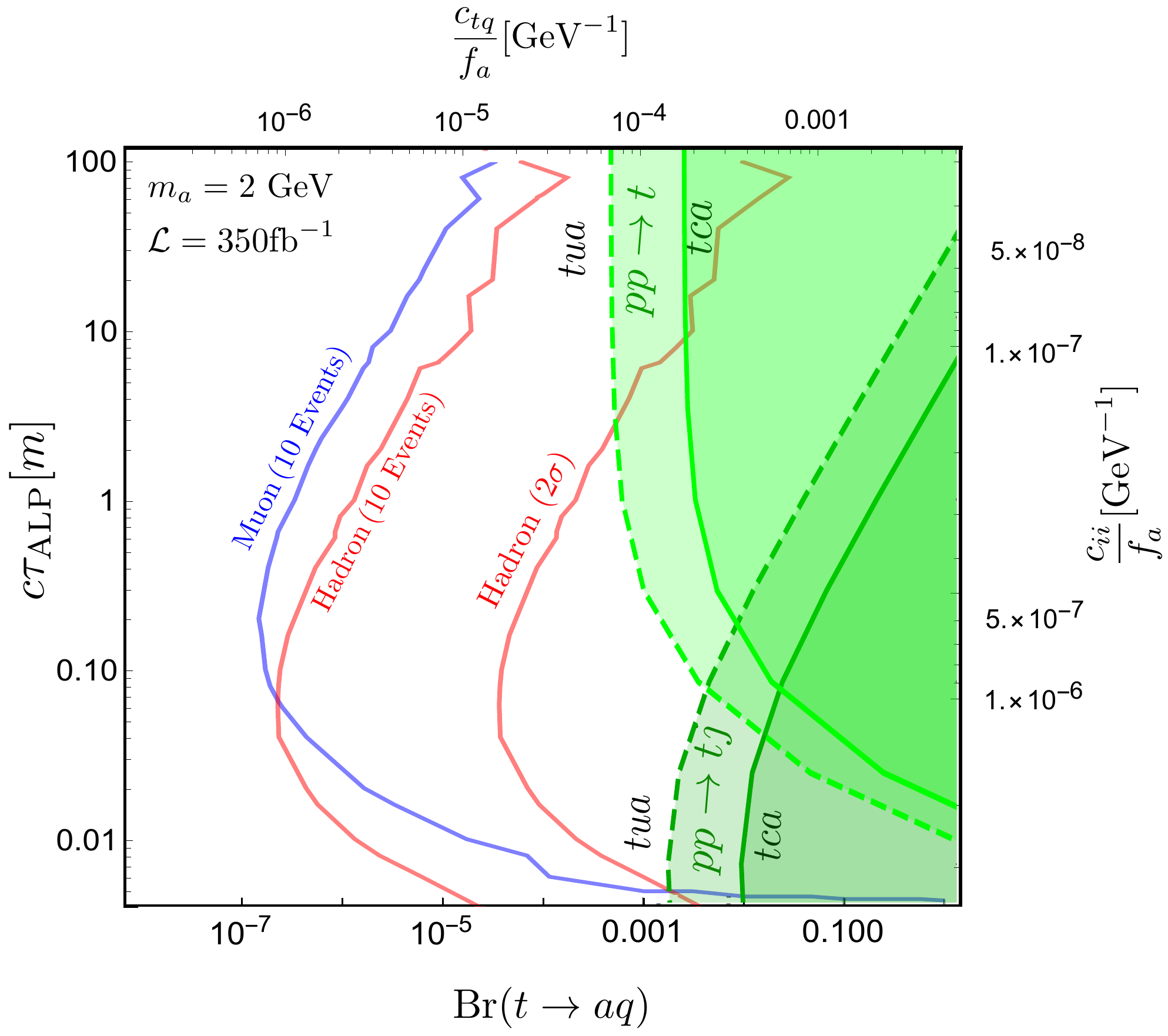}\\
    \quad
    \includegraphics[width=0.64\linewidth]{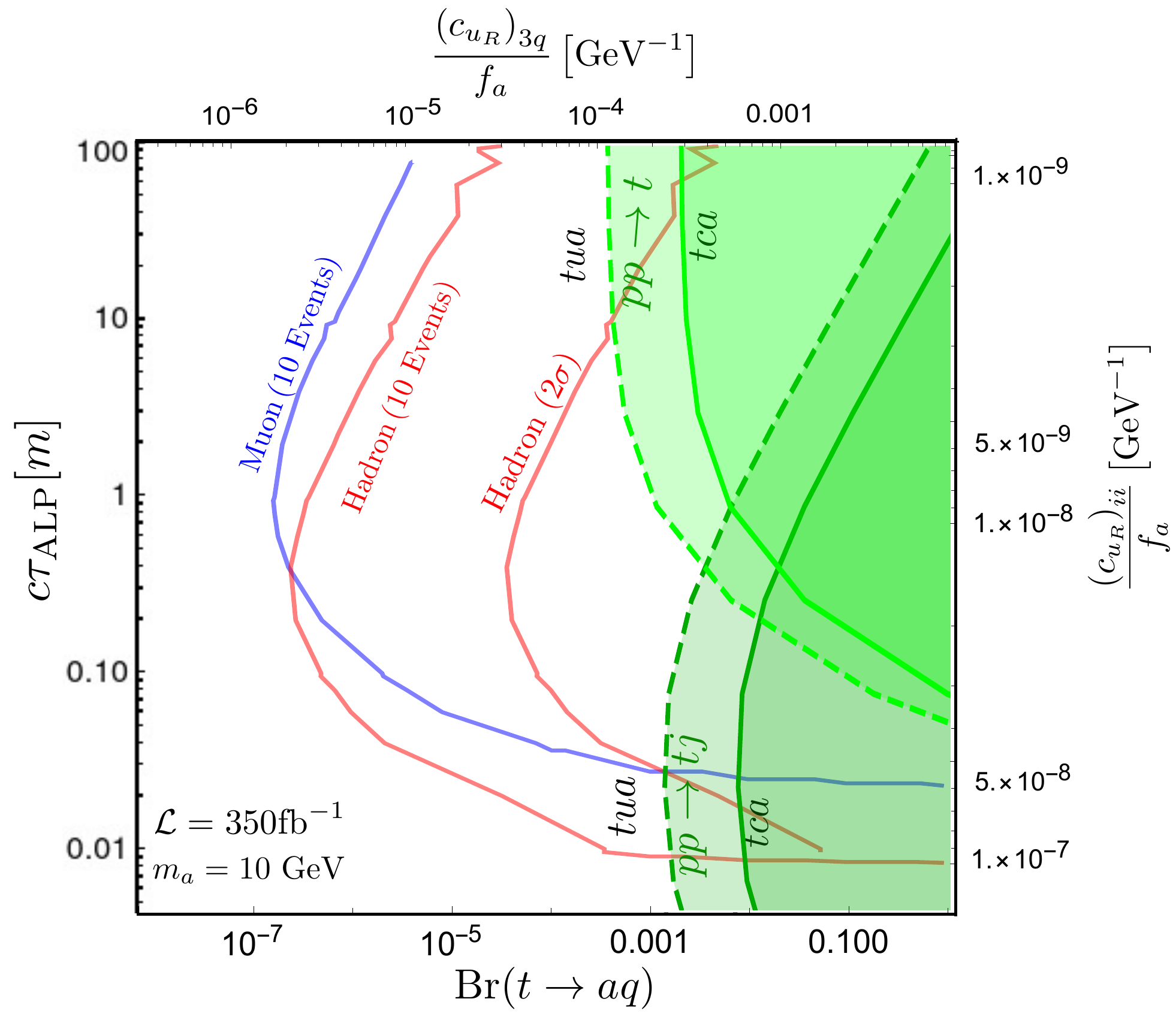}
    \caption{The expected bounds as a function of lifetime $(c\tau_{\rm ALP})$ and the branching ratio of the exotic top decay $\text{Br}(t\to aq)$, for $m_a = 2 \ \text{GeV}$ (top) and $m_a = 10 \ \text{GeV}$ (bottom). The red line labeled with `Hadron 2$\sigma$' represents the conservative limit (Tab.~\ref{table:cutflow}) where $\frac{S}{\sqrt{S+B}} =2$, assuming $\mathcal{L} = 350 \ \text{fb}^{-1}$ integrated luminosity. The red (blue) solid line is the potential discovery line where 10 signal events are produced in the hadronic (muon) calorimeter, in case a background free search can be designed. Finally, the green shaded regions indicate the current bounds on the model. The dark green lines are derived from the top + jet~\cite{CMS:2016uzc}  final state, and the light green lines are from the single leptonic top search~\cite{Goldouzian:2016mrt,Goldouzian:2016ufu}. The dashed lines are for the constraints on the $tua$ coupling, and the solid ones are that of the $tca$ coupling.}
    \label{fig:exclusion}
\end{figure}

In Fig.~\ref{fig:exclusion} we show the expected $2\sigma$ exclusion region of the here proposed search for $\sqrt{s} = 13$~TeV and $\mathcal{L}= 350$~fb$^{-1}$ as the red solid line (labeled with `Hadron (2$\sigma$)').  The bounds discussed in Section~\ref{sec:recast} from recasting the top + jet and single top + missing energy searches are displayed as dark and light green shaded regions, respectively. The regions inside the dashed lines show the constraints from the $tua$ coupling and the regions inside the solid lines from $tca$ coupling.  Finally we show 
the $10$-event discovery lines~\footnote{10 events were chosen to leave
some room for a loss of signal efficiency, and in order to remain conservative.} for the above discussed search in the muon system (blue line) and for a background free search in the hadronic calorimeter (red line), to highlight the potential reach of further improved searches. In the upper panel we use $m_a = 2$~GeV and  in the lower panel $m_a = 10$~GeV. 

For small ALP lifetimes $c\tau_{\rm ALP}\lesssim 0.006$~m ($c\tau_{\rm ALP}\lesssim 0.02$~m) the top + jet search is the the most sensitive constraint and excludes branching ratios down to $\mathrm{Br}(t\to a q)\sim 0.001$ for $m_a = 2$~GeV ($m_a = 10$~GeV). Top + jet searches can probe the exotic top ALP coupling up to $c\tau_{\rm ALP}\sim 30$~m ($c\tau_{\rm ALP}\sim 100$~m) for large enough branching ratios. These bounds arise for ALPs decaying inside $2.5$~cm $< r < 2$~m. For $m_a = 10$~GeV in the lower panel one can also see the exclusion line for ALPs decaying before $0.01$~cm in the lower right corner. On the other hand, the single top + missing energy search only becomes sensitive for $c\tau_{\rm ALP} \gtrsim 0.01$~m ($c\tau_{\rm ALP} \gtrsim 0.1$~m) and is more sensitive than our newly proposed search for $c\tau_{\rm ALP} \gtrsim 1$~m ($c\tau_{\rm ALP} \gtrsim 10$~m). In this region $\mathrm{Br}(t\to a q) \gtrsim 10^{-4}$ is excluded for both masses. Single top searches leave the intermediate lifetime region ($c\tau_{\rm ALP}\sim 0.006-1$~m for $m_a = 2$~GeV and $c\tau_{\rm ALP}\sim 0.02-10$~m for $m_a = 10$~GeV) largely unconstrained. The here proposed search is sensitive in this region as shown in Fig.~\ref{fig:exclusion}. For both $m_a = 2$~GeV and $m_a = 10$~GeV exotic top decays with branching ratios smaller than $\mathrm{Br}(t\to a q) =10^{-4}$ can be probed with $2\sigma$ significance by using the Calorimeter Ratio trigger requirements as event selection criteria. Different masses influence at which lifetimes this search reaches its highest sensitivity, since ALPs with larger masses are less boosted. Here, for $m_a=2$ $(10)$~GeV the search is most sensitive at $c\tau_{\rm ALP} \sim 0.04$ $(0.3)$~m. 

Finally we assume that a more advanced search strategy for long lived ALPs from exotic top decays could be made virtually background free, e.g. by exploiting the differences in the calorimeter showers between signal and background. The $10$-event discovery lines for such a search and for a similarly background free search in the muon spectrometer suggest that probing the intermediate lifetime regime down to branching ratios as small as $\mathrm{Br}(t\to a q)\sim10^{-7}$ is possible.

Fig.~\ref{fig:exclusionHL} in the Appendix shows additionally the expected discovery lines for background free searches for decays in the hadronic calorimeter and the muon system for the HL-LHC. There branching ratios as low as $10^{-8}$ can be reached. Note that the same $p_T$ requirement as for LHC have been used for the jets and the ALPs. Optimizing them for HL-LHC could move the expected $10$-event lines to even smaller branching ratios.

\section{Conclusions}

In this work we have presented a new search for long-lived ALPs  with flavor-violating couplings to up-type quarks involving exotic top decays in top-pair production. We concentrated on topologies where one of the top quarks decay to an ALP together with an up- or charm-quark, with the ALP subsequently decaying in the hadronic calorimeter, leading thus to a large value of the hadronic to electromagnetic energy ratio $E_{\rm had} /E_{\rm  em}$. We have demonstrated that  a cut on this ratio, together with track vetoes, are enough to suppress the $t\bar{t}$ background to the point of being able to test exotic top branching ratios below $10^{-4}$ for $m_a\sim\mathcal{O}(1)$\,GeV in the next run of LHC. We also studied the potential reach of more refined searches and show projections for the high luminosity LHC. Moreover, we have presented a recast of existing single top searches and derived new constraints for both prompt ALP decays as well as for detector stable ALPs. Together these searches can probe exotic decays of the top quark to ALPs across the full range of ALP lifetimes. Our newly proposed search here can increase the sensitivity by more than an order of magnitude for ALP lifetimes in the centimeter to meter range. 

Given the large number of $t\bar{t}$ events at the LHC, further improvements of the search strategy might be possible. In particular here we have not used the shape of the shower in the calorimeters, which could provide further discrimination between the signal and background: an ALP decaying inside the hadronic calorimeter should look quite different from a jet that travels through the full calorimeter. Due to the small ALP mass and large boost factor, the jets should also be unusually narrow. Similarly we believe that a search in the muon system could be essentially background free. In both cases, as few as 10 events might be enough to observe this exotic top decay, and thus probe branching ratios as small as $10^{-7}$. 

There are other potentially interesting signatures which we have not discussed here. Once the ALP is embedded again in a more UV complete theory such as the dark QCD scenario, the top quark could decay into an emerging jet, a spectacular signature which should easily stand out. Furthermore it would then be interesting to connect these experimental signatures with the phenomenology of dark matter in such models. We plan to address some of these exciting possibilities in the future.

\begin{acknowledgments}
We would like to thank R. Goldouzian, and H. Mehrabpour for useful discussions. AC acknowledges funding from the European Union’s Horizon 2020 research and innovation programme under the Marie Sk\l{}odowska-Curie grant agreement No 754446 and UGR Research and Knowledge Transfer Found - Athenea3i. This work has been also partially supported by the Ministry of Science and Innovation and SRA (10.13039/501100011033) under grant PID2019-106087GB-C22 and by the Junta de Andaluc\'ia grant A-FQM-472-UGR20. Work in Mainz was supported by the Cluster of Excellence Precision Physics, Fundamental Interactions, and Structure of Matter (PRISMA+ EXC 2118/1) funded by the German Research Foundation(DFG) within the German Excellence Strategy (Project ID 39083149), and by grant 05H18UMCA1 of the German Federal Ministry for Education and Research (BMBF).

\end{acknowledgments}

\clearpage

\appendix
\section{Sensitivity at the High Luminosity LHC}
\label{app:HL}
\begin{figure}[!h] 
    \centering
    \includegraphics[width=0.53\linewidth]{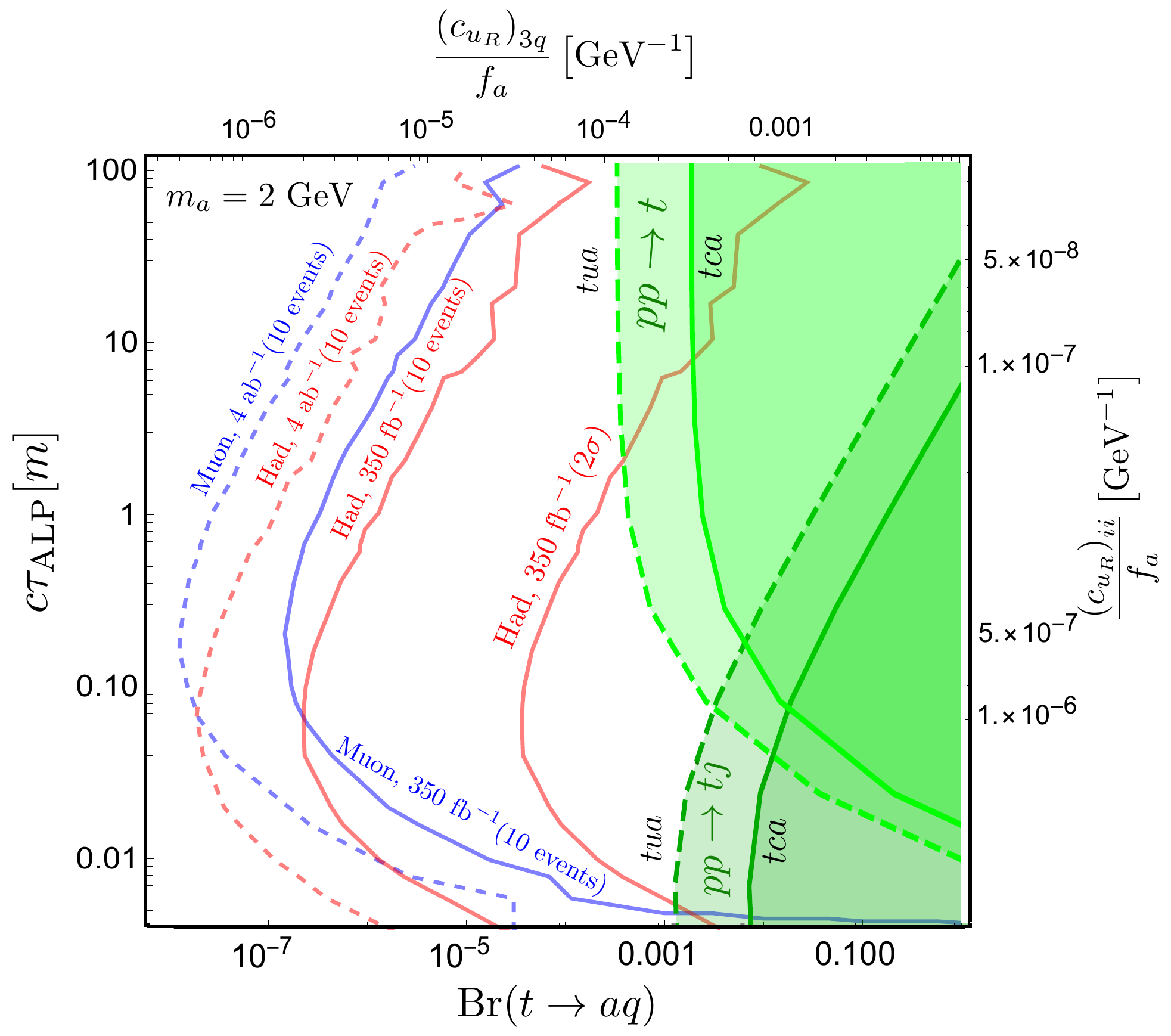}\\
    \quad
    \includegraphics[width=0.53\linewidth]{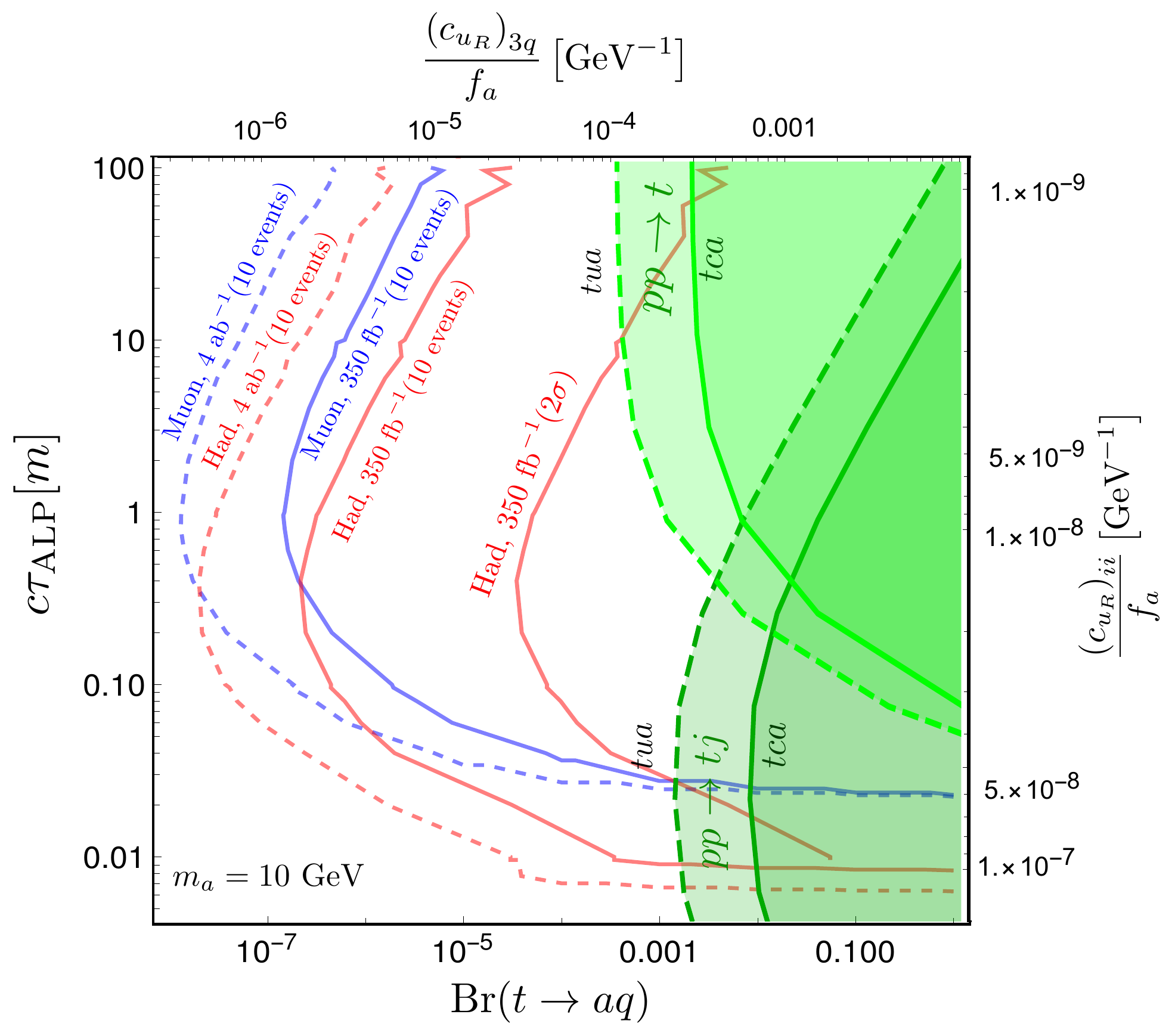}
    \caption{The same figure as Fig.~\ref{fig:exclusion}, where the expected 10 events lines at the HL-LHC with $\sqrt{s} = 14 \ \text{TeV}$, and $\mathcal{L} = 4 \ \text{ab}^{-1}$ integrated luminosity is shown in dashed lines. The red dashed line represents an ALP that decays within the hadronic calorimeter and the dashed blue is when the ALP decays in within the muon calorimeter.}
    \label{fig:exclusionHL}
\end{figure}

The projected sensitivity at HL-LHC ($\sqrt{s} = 14 \text{TeV}$, and integrated luminosity $\mathcal{L} = 4 \text{ab}^{-1}$) is shown with dashed lines in Fig.~\ref{fig:exclusionHL}. These lines indicate the potential discovery requiring 10 signal events using the same cuts as the current run of the LHC, and assuming no backgrounds. With the higher luminosity, and the increase in the pile-up effect, the dashed lines may need to be adjusted.
\section{Cut flow for three to five jets}
The ALP in this search is highly boosted and decays displaced. Thus, while it decays to a pair of partons, it will be seen mostly as one narrow jet. In this case the signal has maximal five final state jets, including the one from the ALP decay. For comparison we show in Tab.~\ref{table:cutflow2} the cut flow (of the efficiency) for changing the cut on the number of jets from $3-6$ to $3-5$ jets. 
\begin{table}[th]
    \centering
    \begin{tabular}{l|c|c|c}
         & $\mathbf{m_a = 2}$~\textbf{GeV} & $\mathbf{m_a = 10}$~\textbf{GeV} & $\mathbf{t\bar{t}}$ \\
         \hline
        \textbf{total} & (1) $2.79\times10^{5}$ & (1) $2.79\times10^{5}$ & (1) $2.91\times 10^{8}$ \\
        \hline
        $\mathbf{3-5}$ \textbf{jets with}  &  & &\\
        \textbf{$\mathbf{p_T>40}$~GeV \&} $\mathbf{|\eta| < 2.5}$ & (0.7815) $2.18\times10^{5}$& (0.7779) $2.17\times10^{5}$ & (0.65997) $1.92\times10^{8}$ \\
        \hline
        \textbf{1 jet with} $\mathbf{log_{10}\left(\frac{E_{\rm had}}{E_{\rm em}}\right)> 1.2}$ & (0.1330) $3.71\times10^4$ & (0.0699) $1.95\times10^4$ &(0.01022) $2.97\times 10^{6}$\\
        \hline
         \textbf{displaced jet has $\mathbf{\leq 2}$ tracks}& (0.1330) $3.71\times10^4$ & (0.0699) $1.95\times10^4$ & (0.00018) $5.23\times 10^4$ \\
         \textbf{with $\mathbf{p_T > 2}$~GeV}& & & \\
    \end{tabular}
    \caption{Cut flow of the expected number of events for signal and background events for LHC run 3 with $\sqrt{s} = 13$~TeV and $\mathcal{L}= 350$~fb$^{-1}$. The values in brackets are the efficiencies after each cut. For the signal $c\tau_{\rm ALP} = 0.1$~m and $\mathrm{Br}(t\to a q) = 0.001$ was chosen.}
     \label{table:cutflow2}    
\end{table}

\bibliography{refs}{}
\bibliographystyle{JHEP}

\end{document}